# Multiplex Imaging Analysis in Pathology: a Comprehensive Review on Analytical Approaches and Digital Toolkits.

Mohamed Omar[a,b,*], Giuseppe Nicolo' Fanelli[a,c,*], Fabio Socciarelli[a], Varun Ullanat[b], Sreekar Reddy Puchala[b], James Wen[b], Alex Chowdhury[a,b], Itzel Valencia[a], Cristian Scatena[c], Luigi Marchionni[a,b], Renato Umeton[a,b,d,e,#], Massimo Loda[a,b,f,#]

[a] Department of Pathology and Laboratory Medicine, Weill Cornell Medicine, New York, NY, US

[b] Dana-Farber Cancer Institute, Boston, MA, US

[c] Department of Translational Research and New Technologies in Medicine and Surgery, University of Pisa, Pisa, IT

[d] Harvard T.H. Chan School of Public Health, Boston, MA, US

[e] Massachusetts Institute of Technology, Cambridge, MA, US

[f] Broad Institute of MIT and Harvard, Cambridge, MA, US

*Co-first authors

#Co-corresponding authors: renato_umeton@dfci.harvard.edu and mloda@med.cornell.edu

## Abstract

Conventional histopathology has traditionally been the cornerstone of disease diagnosis, relying on qualitative or semi-quantitative visual inspection of tissue sections to detect pathological changes. Immunohistochemistry, while effective in detecting specific biomarkers, is often limited by its single-marker approach, which constrains its ability to capture the full complexity of tissue environment. The introduction of multiplexed imaging technologies, such as multiplexed immunofluorescence and spatial transcriptomics, has been game-changing by enabling the simultaneous visualization of multiple biomarkers within a single tissue section. These advanced techniques complement morphological data with quantitative molecular and spatial information, offering a more comprehensive view of the tissue microenvironment, cellular interactions, and disease mechanisms. This multimodal perspective is critical for understanding disease progression, patient prognosis, and treatments response. However, the complexity and scale of data generated by multiplexed imaging present significant challenges. The vast amount of data produced by multiplexed imaging requires complex computational methods for preprocessing, segmentation, feature extraction, and spatial analysis. These tools are essential for handling large, multidimensional datasets and converting raw imaging data into actionable insights. By automating labor-intensive tasks and improving the reproducibility and accuracy of results, computational tools have become pivotal in both research and clinical diagnostics. This review presents a comprehensive overview of the current landscape of multiplexed imaging in pathology, detailing the entire workflow—from selecting the most appropriate multiparametric imaging modality to its downstream applications. It also highlights key technologies and data processing techniques that facilitate the integration of advanced methods into routine pathology practice through an end-to-end workflow, such as PathML, an AI-powered platform designed to streamline multiplexed image analysis, offering a modular, user-friendly solution that simplifies the analysis and interpretation of complex datasets, making it accessible for everyday clinical and research use.

# Contents

## 1. Introduction

Pathology has long been a cornerstone for detecting, analyzing, and understanding disease processes, bridging basic science and clinical practice. Traditionally, pathology focuses on the study of tissues, cells, and bodily fluids. Conventional histopathologic techniques and staining such as Hematoxylin and Eosin (H&E), are designed for the evaluation of cellular morphology and tissue architecture for diagnostic purposes. However, the inherently qualitative nature of these examinations highlights the need for more quantitative, standardized approaches that can leverage the wealth of information available in whole slide images (WSIs) [1,2].

One of the first revolutions in pathology has been immunohistochemistry (IHC), which is currently used to visualize and quantify the expression of specific diagnostic, prognostic, and predictive biomarkers directly on tissue samples. In routine diagnostics, IHC is typically used to assess a single biomarker per tissue section requiring pathologists to examine multiple slides sequentially. This becomes cumbersome and challenging, particularly for undifferentiated tumors, cancers of unknown primary, sarcomas, and most lymphoid neoplasms which require the evaluation of multiple biomarkers and their co-expression to be fully characterized [3]. Currently, flow cytometry is the only method available in most clinical laboratories for multiantigen labeling of individual cells. Still, this method is unsuitable for many sample types, including formalin-fixed paraffin-embedded (FFPE) tissues, and lacks spatial information [4]. Moreover, for tumors where sampling is limited to fine needle aspirates (FNA) or core needle biopsies (CNB), using multiple sections for singleplex IHC and splitting the biopsy for molecular testing can compromise diagnostic accuracy. This issue is particularly relevant for advanced cancer patients, who often cannot undergo surgery and diagnostic, prognostic, and predictive information must be obtained only from FNA or CNB. For such cases, multiplex staining allows the assessment of all relevant biomarkers of interest using limited tissue sections [3].

Unlike traditional imaging that provides a singular, often morphological or semiquantitative perspective, modern multiplexed imaging techniques, such as multiplexed immunofluorescence (mIF) or immunohistochemistry (mIHC), and high-plex immunofluorescence/imaging mass cytometry offer a comprehensive view of the tissue microenvironment by mapping the expressions of several protein markers in their native histological context [4–10]. This multiplexed approach enables the simultaneous assessment of multiple biomarkers, offering a more comprehensive view of the disease state that complements the insights gained from traditional H&E staining alone. The integration of such diverse data through computational analysis promises not only to enhance the accuracy of existing predictive and prognostic models but also to uncover novel insights into disease mechanisms. For instance, by simultaneously visualizing and quantifying the expression of multiple biomarkers within a single tissue section, pathologists can obtain a nuanced understanding of tumor biology including, but not limited to, cellular composition, spatial interactions between different cell types, and patterns of immune infiltration. Investigating these dynamics across different disease stages or different treatment modalities can in return identify features or biomarkers associated with cancer progression or treatment resistance.

Despite the advances brought by multiplexing technologies in quantitative histopathology research, transitioning to a computational analysis-driven approach involves significant challenges. The sheer volume of data produced by multiplexed imaging requires extensive computational resources and sophisticated algorithms for tasks such as preprocessing, cell detection, and tissue or cellular features quantification. These processes are crucial for translating raw pixel data into actionable single-cell information that can be analyzed quantitatively [11]. However, beyond these technical hurdles, lies substantial potential to aid both pathology practice and biomedical research, especially in fields like oncologic pathology. To fully realize this potential, a multidisciplinary approach is essential. Effective integration of skills from pathology, computer science, and medical research is required to develop and refine computational tools that can manage and interpret the complexity of multiplexed data. This interdisciplinary collaboration is not merely a necessity but an opportunity to synergize knowledge across fields, driving forward the capabilities of digital pathology.

Building on this foundation, this review outlines a spectrum of tasks pertinent to the computational analysis of histopathology multiplexed imaging (**Figure 1**). We explore practical computational tools that, at times leveraging artificial intelligence (AI), streamline these complex tasks, ranging from preprocessing and quality control—including primary antibody validation and spectral unmixing—to advanced computational algorithms for nuclei and cell segmentation. The subsequent discussion delves into feature extraction, quantification techniques, and methods for automated cell annotation, employing both gating-based and unsupervised clustering-based approaches. Given the complexity of such analysis and the plethora of tools involved, we provide guidance on end-to-end workflows for multiplexed image analysis. We conclude with an introduction to PathML v3, a modular workflow designed for advanced multiplexed image analysis, and "Talk to PathML", a digital pathology assistant aimed at democratizing advanced computational image analysis for pathologists and researchers with limited coding experience [12]. Before proceeding, take a moment to consult ***Supplementary Table 1*** for an overview of the key terms and concepts used in this paper.

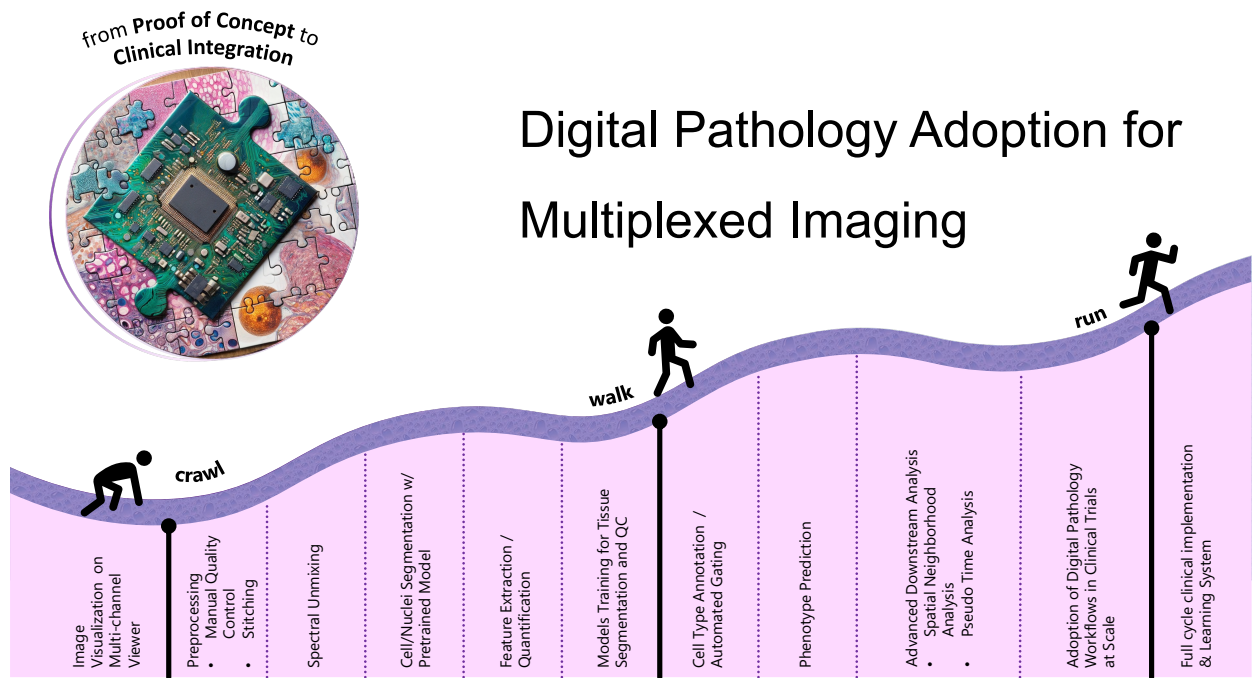

**Figure 1.** Overview of common pathology tasks and digital tool implementation for the analysis of histopathology multiplexed imaging data. This figure illustrates the range of tasks commonly encountered in the computational analysis of this data, categorized by their relative ease of implementation.

## 2. Multiparametric Imaging Modalities: One Size Does Not Fit All

Histological imaging has seen an incredible increase in acquisition modalities associated with IHC in the last ten years. Once limited to one or two markers and mostly relegated to chromogenic detection, IHC is now experiencing a substantial expansion, in multiplexing capacity and detection techniques, with innovative technologies allowing for the detection of up to 40 markers on a single slide [12,13].

Keeping track of the myriad multiplexing technologies and detection techniques is a formidable challenge due to their rapid evolution. Therefore, we will describe the main approaches present in pathology research. **Table 1** provides a comparative overview of the various multiplexed imaging modalities present in the literature.

While our review mainly focuses on protein-based markers with subcellular-level resolution in 2D formats, it is important to acknowledge additional imaging modalities. Noteworthy among these are mass spectrometry imaging, which facilitates the detection of lipids, metabolites, drugs [13], various forms of spatial transcriptomics, ranging from 10X Genomics Visium technology [14] to NanoString GeoMx ROI-based sampling [15], multiplex FISH [16], and 3D clearing microscopy [17].

Each of these techniques contributes unique insights into tissue structure and function, expanding the frontiers of histological analysis.

**Table 1.** Imaging modalities for multiplex staining.

| | Brightfield multiplex IHC[18,19] | TSA multiplex IF[20] | Cycling mIF and stripping[9,21] | Imaging Mass Cytometry[22–24] |
|---|---|---|---|---|
| *Detection system* | Chromogens with narrow absorbance band and dedicated camera | Tyramide-conjugated fluorophores associated with HRP-conjugated secondary ab | Fluorophore-conjugated secondary antibodies, imaging and following stripping | Primary antibodies conjugated with metals |
| *Image dimensions* | 2D | 2D, 3D | 2D, 3D | 2D |
| *Channel multiplexing* | Low (6) | Low (max 8) | Very high (up to 100) | High (up to 40) |
| *Key Advantages* | High resolution, not expensive, can be observed at the brightfield microscope | Robust protocol, very sensitive, fast | High-plex, good resolution, very well automated | Fast unmixing, useful if N fluorophores > N detection channels |
| *Potential pitfalls* | Chromogens not resistant to time, need for a dedicated camera, low-plex | Traumatic for the tissue, low-plex, preservation of antigenicity | Not always sensitive, lengthy protocol, expensive, some approaches have short expiration date reagents, preservation of antigenicity after repeated cycle | Very long protocol, small areas acquired, very expensive, low resolution, traumatic for the tissue, antibody conjugation an issue |
| *Commercial solutions* | Roche Diagnostics, Enzo Life Sciences, Leica Biosystems | Akoya Biosciences, Thermo Scientific, Biotium | Lunaphore, Akoya biosciences, Cell Signaling Technology, Ultivue | Standard BioTools, IonPath |

## 3. Primary Antibody Validation: The Right Thing at The Right Place

After the choice of the proper multiplex imaging technique, the second step is the careful validation of each biomarker performance, singularly and in the context of a multiplex panel.

Addressing the complexities of antibody validation is a task often underestimated. Numerous instances of the detrimental consequences of using poorly validated reagents underscore the importance of rigorous performance checks before their use [25,26]. While this review primarily focuses on the analytical workflow for multiplexed images, it is also crucial to address the experimental work that precedes image creation, which can substantially impact image quality and subsequent analyses. Singleplex IHC assays are designed and often optimized directly by vendors to detect the presence or absence of specific proteins. In contrast, multiplex assays need strong validation to use the wide range of protein expression levels, provide continuous measurements, and integrate spatial information. The design and validation of multiplex panels require expertise in immunology, histopathology, and tissue staining techniques [27].

Although robust standard operating procedures and commercial solutions are readily available for various imaging modalities, pre-analytical variables, and primary antibody validation remain a critical bottleneck despite the presence of numerous guidelines [28–33]. Cold ischemia time should be minimized to preserve enzyme activity, protein integrity, cytomorphology, and prevent autolysis. Fixation time and extent must adhere to strict recommendations, as delayed fixation prolongs ischemic time and overfixation can harm antigenicity. Additional critical variables include tissue processing steps, slide-drying, and storage conditions [34].

When selecting antibodies, monoclonal antibodies are preferred over polyclonal due to higher specificity and consistency [35], and notable initiatives like Human Protein Atlas [36], HubMap [37], Genecards [38], and others [39] can be helpful. Well-validated and reproducible antibodies, along with properly validated controls, are widely accessible. However, many other targets lack readily available antibodies, needing extensive validation using different tissue and techniques such as western blotting of control cell lines [29]. Optimizing multiplex assays requires figuring out the best antibody dilution, incubation time, and staining order. This starts with singleplex IHC for each antibody, then each marker is paired with a suitable chromogen/fluorophore, considering its expression level and its characteristics [34]. Upon completing panel optimization, each stain should perform equivalently to singleplex IHC in terms of control performance, expression intensity, and cell-type-specific labeling [40]. Troubleshooting may involve adjusting staining order, antibody concentration, fluorophore concentration, or fluorophore-marker pairing to address issues like steric hindrance or signal overlap [35]. Finally, image acquisition and analysis also require thorough validation, with standardized field selection to avoid bias and comparison with H&E-stained slides at critical steps.

## 4. Fluorophores and Image Acquisition: Illuminating the Pathways

In immunofluorescence, signal detection is made by acquiring different wavelengths across the visible light and infrared spectrum. Multiple specialized filters and fluorophores, each designed to emit light at specific wavelengths when excited, are available for different uses.

Employing multiple fluorophores with non/partial-overlapping emission spectra allows for the concurrent detection and differentiation of multiple targets within a single sample, providing a detailed and comprehensive view of the biological processes and interactions occurring in real-timer [41].

Each fluorophore used in fluorescent imaging has its own set of advantages and limitations, primarily defined by the wavelength at which it emits light. Understanding these characteristics is crucial for selecting the proper one for specific imaging needs and improving the quality and accuracy of the data obtained.

Fluorophores emitting in the blue-to-green spectral ranges (*e.g.*, DAPI, GFP) have shorter wavelengths, which provide higher spatial resolution due to their smaller diffraction limit. This makes them particularly effective for detecting fine details and observing smaller structures within cells, such as nuclei or cytoskeletal elements. For fixed samples, these fluorophores are ideal as they allow for detailed imaging at a cellular and even subcellular level. However, these shorter wavelengths carry higher energy, which can be problematic in 'live/fresh' imaging applications. High-energy light can cause photobleaching, where the fluorophore loses its ability to fluoresce after prolonged exposure, leading to signal loss over time. Furthermore, these fluorophores are more likely to cause photodamage to the cells or tissues being observed, potentially altering biological processes or causing cell death. Therefore, while short-wavelength fluorophores offer precision, they must be used cautiously, particularly in experiments involving live samples or repeated imaging [42].

Fluorophores emitting in the red or near-infrared (NIR) regions (*e.g.*, Cy5, Alexa Fluor 647) have longer wavelengths and several distinct features. These fluorophores are associated with lower energy levels, making them less likely to cause photodamage or photobleaching, thus extending their utility in long-term or live imaging experiments. Additionally, longer wavelengths penetrate deeper into tissues compared to shorter wavelengths, making them highly suitable for imaging thick tissue samples, 3D cultures, or *in vivo* studies where deeper tissue structures must be visualized. This deep penetration is particularly helpful for applications such as whole-organ imaging, brain imaging in live animals, or imaging through complex tissue layers like skin. However, there are trade-offs. The longer wavelengths used by these fluorophores inherently produce lower-resolution images because of the larger diffraction limit, making it challenging to capture fine cellular details with the same precision as short-wavelength fluorophores. Additionally, the fluorescence signals from these wavelengths can be weaker, requiring the use of more sensitive detectors or more powerful light sources to achieve adequate signal strength. This can add complexity and cost to the imaging setup, and in some cases, may still result in a lower signal-to-noise ratio compared to shorter-wavelength options. Moreover, certain biological tissues can absorb or scatter light at longer wavelengths, which may reduce image clarity or contrast if not properly accounted for during imaging [43].

Overall, the choice between short- and long-wavelength fluorophores depends on the specific application requirements, including the depth of imaging needed, the resolution desired, and the nature of the sample (fresh or fixed). Combining fluorophores across the spectrum, while managing their specific limitations, can optimize imaging performance and provide comprehensive insights into complex biological structures and processes. However, exposure times need to be set up carefully to maintain a balance of the signal intensity across markers in the panel.

Finally, autofluorescence should be considered during image acquisition. It refers to the natural emission of light by certain tissues or cellular components, such as collagen or NADH, which can interfere with fluorophores detection. Autofluorescence can create background noise, reducing the clarity and contrast of the images. To mitigate this, careful selection of fluorophores with emission wavelengths distinct from those typically associated with autofluorescent molecules is essential. Advanced filtering, spectral unmixing (see below), and processing techniques are also employed to minimize the impact of autofluorescence, enhancing the overall image quality [27].

After the staining, slides need to be acquired and when opt for a scanner system for image acquisition, several factors must be considered, including the spectral range, fluorescence throughput, automation features, multiplexing capabilities and camera resolution, among others, to ensure the capture of high-quality images. PhenoImager HT™, one of the most used multispectral digital slide imaging systems, employs proprietary multispectral imaging technology to mitigate optical spectral bleed-through between channels and effectively distinguish signal from background autofluorescence. In an internal evaluation, the average optical bleed-through was 8.7% for a 6-plex assay and 13% for an 8-plex assay and multispectral unmixing reduced residual bleed-through to less than 1% in both cases [40].

## 5.  Spectral Unmixing: Maximizing Separation for Clear Insights

Spectral unmixing is a fundamental technique widely used in confocal microscopy to separate fluorescent signals within 'lambda stacks'— which involve capturing a broad range of spectra, typically far exceeding the number of fluorophores used in the sample. This approach allows for the differentiation of overlapping emission spectra, providing a clear separation of signals even when multiple fluorophores are present. However, in multiplex TSA-based imaging technique, the spectral unmixing approach is different: the number of spectra acquired by the camera exactly matches the number of fluorophores being detected. This precise correlation optimizes signal detection and minimizes spectral overlap, which is critical when using TSA to achieve high sensitivity and specificity in detecting multiple targets within the same tissue section [44]. By ensuring that the captured spectra align with the fluorophores used, this approach enhances the accuracy and efficiency of multiplex imaging. However, even with this approach a partial overlap of absorption and emission spectra among the fluorophores used is still present, leading to channel bleed-through [45]. To address this challenge, various computational algorithms have been developed and the most used are summarized below [46].

Linear unmixing (or linear decomposition) is the most common algorithm used for spectral unmixing. This method can calculate the different contributions of each fluorophore to every channel of the image using as a reference a library of fluorescent spectra acquired singularly [30]. In case an autofluorescence-dedicated channel is acquired, the linear unmixing is also able to eliminate the autofluorescence from the tissue.

PICASSO is an algorithm that minimizes the "mutual information" by subtracting iteratively one scaled channel image from another [47]. This unmixing approach has been used on multiplex TSA immunofluorescence with success from the authors, reporting a superior performance of this algorithm than linear unmixing.

Maric *et al.* developed an algorithm based on linear unmixing in which the bleed-through between channels is estimated using the LASSO regression and a semi-supervised model is used to separate the different channels [48]. Although library-free, this technique requires the user to indicate which couples of channels are expected to show bleed-through.

LUMoS [49] is an algorithm based on clustering machine learning and has been developed for images in which the number of fluorophores is equal or superior to the number of detectors. The drawback of this method is that it has only been used on two-photon microscopy and never applied to widefield fluorescent microscopy, limiting its application in the pathology field.

Finally, an algorithm based on non-negative matrix factorization (NMF) [50] has been applied to widefield microscopy on tissue but its efficacy has been questioned by other authors [51,52].

Key advantages, potential pitfalls, and several other aspects of these algorithms are detailed in ***Supplementary Table 2.***

## 6. Image Preprocessing: Optimizing Data for Accurate Analysis

Image preprocessing involves several essential techniques to prepare images for the following steps of segmentation, feature extraction, and quantification. Resizing/tiling ensures images are of a uniform size, which is crucial for the effective functioning of machine learning algorithms. Grayscaling simplifies the image data by converting color images to grayscale, reducing computational requirements for certain algorithms. Binarization converts grayscale images to black and white through thresholding, while contrast enhancement uses methods like histogram equalization to improve image clarity. To remove unwanted noise, techniques such as smoothing, blurring, and filtering are applied. Normalization adjusts pixel intensity values to a specific range, typically between 0 and 1, to enhance the performance of models. By applying the right combination of these techniques, you can significantly enhance your image data, leading to the development of more effective computer vision applications. These preprocessing steps are critical components of the image analysis pipeline, proving a solid foundation for achieving accurate and reliable outcomes.

## 7. Quality Control: Precision and Accuracy at Every Step

For the technical performance of the assay, quality control (QC) metrics like those used in singleplex IHC, such as batch-to-batch differences in antibody performance and antibody performance in decalcified specimens, must be considered [35]. Additionally, scan quality variability should be minimized [40].

Quality assessment of slides is crucial to prevent the incorporation of low-quality data into the analytical pipelines. Common artifacts such as tissue folding, air bubbles, out-of-focus areas, tissue detachment, foreign bodies, and poor staining quality are prevalent in virtual slides[40] and can significantly undermine downstream analysis [53,54]. For brightfield and singleplex digital slides, several digital and automated QC tools or pipelines are readily available [55], however, for multiplexed images, these options are limited, and manual quality checks by experienced pathologists remain the gold standard [56]. Among the few published methods, the most promising are listed below.

Jiang *et al.* [57] created a tool based on the DAPI channel signal able to recognize some artifacts like blurring, foreign bodies, halo artifacts, and folding.

StainV&QC, a plugin tool for TissuUmaps, is used to verify the staining quality. While this tool requires a preliminary step of cell segmentation, it provides an important evaluation of the staining conditions and background artifacts [58].

MxIF Q-score [59], is a tool able to evaluate the quality of image registration, TMA core quality, and cell segmentation on fluorescent digital slides. The algorithm and metrics used are completely based on DAPI fluorescence.

Finally, CellProfiler is a noteworthy widely used image analysis software primarily designed for cell-based immunofluorescence [60]. One of its available modules ("MeasureImageQuality") measures several metrics of quality, mostly related to blurring but also correlated to saturation of signal (often the result of folding or out-of-focus areas) and too low estimation of the exposure time. In our experience, certain blurring metrics were dependent on the cellular content present in the images, indicating that variations in cell density or structure could impact the results. This observation aligns with guidance from the CellProfiler developers, who emphasize the importance of selecting an appropriate spatial scale when using the power log-log slope method [32]. Choosing the correct spatial scale is critical, as it ensures the metric accurately reflects image quality without being disproportionately affected by the variability in cellular features. This consideration is particularly important when applying automated image analysis techniques to heterogeneous samples, where differences in cell morphology or distribution may otherwise confound the interpretation of image-blurring metrics. Finally, another limitation of this package is that it is not able to read large format images, needing a preliminary tiling step.

## 8. Cell Segmentation and Annotation: Accurate Boundaries for In-Depth Analysis

After all the above steps, to extract meaningful insights from multiplexed images, we need to isolate individual cells from the complex arrangement of tissue samples (cell segmentation) and categorize them based on their unique phenotypic expressions (cell annotation). The reliability of quantitative analyses of cell morphology, spatial distribution, and the relationships between cellular components and their molecular expressions hinges on the accuracy of the preceding cell segmentation.

### 8.1 Advanced Computational Algorithms for Automated Cell Segmentation in Multiplexed Imaging

Cell segmentation techniques range from basic thresholding methods to sophisticated machine learning algorithms. Traditional methods like thresholding and edge detection were initial tools used to distinguish cells from the background based on intensity values or gradients. While these methods are straightforward and computationally efficient, their effectiveness is reduced in heterogeneous tissue samples and variable staining quality. In multiplexed images, where multiple markers are visualized simultaneously, the overlap of fluorescent signals can complicate the segmentation process, making these traditional techniques less suitable for complex analyses. Currently, most cell segmentation methods begin by identifying individual nuclei, typically using a nuclear marker such as DAPI [4]. Following nuclei segmentation, cell boundaries are delineated either by detecting the cytoplasm within a specified radius around each nucleus or, more accurately, by using a cytoplasmic or cell membrane marker [4,56]. Here we summarize the latest array of techniques employed in cell segmentation, highlighting their applications, advantages, and limitations within the context of multiplexed image analysis (**Table 2**).

**Table 2.** List of popular cell and nucleus segmentation tools. GUI: graphic user interface

|  | Mesmer [61] | Cellpose [62] | StarDist [63,64] | CellSeg [65] | Ilastik [66] | UnMICST [67] |
|---|---|---|---|---|---|---|
| ***Algorithm/Method*** | Deep learning-based, with ResNet50 backbone and Feature Pyramid Network | Deep learning-based, utilizing U-Net architecture | Deep learning, with a U-Net-like architecture designed for star-convex shape prediction | Deep Learning, utilizing R-CNN architecture | Machine learning-based, utilizing interactive learning and classification algorithms | Deep learning, utilizing a suite of CNN architectures with real augmentation (intentionally defocused and over-saturated images) |
| ***Input Image Type*** | 2D and 3D | 2D (extended to 3D but without 3D training data) | 2D and 3D | 2D and 3D | 2D and 3D | 2D and 3D |
| ***Training Data*** | Trained using TissueNet, a comprehensive image dataset featuring more than one million paired whole-cell and nuclear annotations from nine organs and | Trained on a diverse dataset comprising over 70,000 segmented objects from a variety of cell images | *A dataset of 497 manually annotated real microscopy images of cell nuclei from the 2018 Data Science Bowl* [68] | *A dataset from the 2018 Kaggle Data Science Bowl, containing 29,464 ground truth segmented nuclei* [68] | Based on sparse user-provided training annotations; no extensive training dataset required | Trained on manually curated data from seven tissue types with ~10,400 nuclei labeled for nuclear contours, centers, and backgroun |
| | captured using six different imaging platforms | | | | | d. Also includes training with defocused and saturated images for real augmentation |
| ***Cytoplasmic Marker Needed*** | yes | Not required; can perform segmentation based on cell morphology and machine learning inference | Optional; model is designed to predict cell shapes, can use cytoplasmic markers for enhanced segmentation | Recommended for optimal segmentation performance, especially in dense tissue samples | Not specifically required; utilizes machine learning on user-defined annotations to segment and classify objects | Not specified |
| ***Usability*** | Model weights can be used in python-based pipelines such as DeepCell; Intuitive web-based interface | Python package with GUI for ease of use, some coding possible for advanced tasks | Python API, with some GUI elements available via plugins or extensions | Primarily command-line interface with some GUI elements | User-friendly GUI designed for biologists or researchers without in-depth computational background | Command-line interface with Python API, may require proficiency in programming and computational image analysis |
| ***Integration*** | Can integrate with common bioimage analysis workflows including PathML [11] | Standalone Python package, integrates with common Python data science tools | Integrates with Fiji/ ImageJ and QuPath ecosystems | Python-based, can be integrated with standard scientific Python stack and image analysis tools like Fiji/ ImageJ | Can be used as a standalone application or integrated with Python or Fiji for automated workflows. Can process data larger than RAM and integrate with existing | *Integrated into MCMICRO* [69] *workflow* (see **Table 4**) |
| | | | | | workflows via command line for batch processing | |
| *Customizability* | Pretrained models available, can be further trained on user-provided datasets | Models can be trained with user data for customized segmentation tasks, designed for continuous improvement by periodically re-training the model using community-contributed data | Pre-trained models available (for 2D only), can be further trained on user-provided datasets | Optimized for fluorescence and brightfield biological microscopy images | User annotations guide the learning process; pre-defined workflows are adaptable to various biological image analysis problems | Not specified |

A significant challenge for segmentation tasks is the scarcity of large, expert-annotated datasets for tissue structures. These datasets are crucial for training models to accurately recognize and classify cellular components in histopathological images. Notably, the process of creating such datasets involves extensive manual annotation by expert pathologists, who identify different tissue and cellular structures within large gigapixel whole slide images (WSIs). Typically, this includes annotating thousands of nuclei or cell types per slide, ideally across various tissue types, and obtained through diverse preparation, preprocessing, and scanning techniques.

While conventional H&E-stained WSIs are commonly available in many institutions, slides prepared for multiplexed imaging modalities are more costly and require specialized equipment and expertise, thereby limiting their availability. As a result, building comprehensive annotation datasets for multiplexed images not only demands substantial financial investment for image generation but also significant resources for manual annotation efforts. Despite these challenges, the availability of annotated multiplexed imaging datasets has been increasing recently (see **Supplementary Table 3**), greatly aiding the development of robust cell segmentation pipelines. These datasets often encompass a variety of multiplexed imaging modalities and tissue types, which is critical for developing models that are robust against technical and biological variability.

## 8.2 Feature Extraction and Quantification

Following cell segmentation, the next critical step before annotation in the analysis pipeline is quantification. This process involves extracting quantitative data from segmented cells or nuclei and translating visual information into numerical values suitable for statistical analysis. Typically, this results in the creation of a feature count matrix, where each row corresponds to a single cell or nucleus, and columns represent various features and metrics of interest. Features often include marker expression levels within individual cells as measured by the fluorescence intensity, which can include the minimum and maximum intensity and the average intensity across the segmented cell area. Additionally, quantification algorithms record the spatial location of each cell, typically recorded as x and y coordinates on the imaging plane, as well as morphometric features such as eccentricity, which measures the elongation of a cell, and size, often reported as the area or volume of a cell or nucleus. Feature extraction methods can be broadly grouped into two main categories: non-deep learning and deep-learning applications.

### 8.2.1 Non-Deep Learning Applications (Pixel- and Cell-Based Applications)

Non-deep learning algorithms for feature quantification primarily use segmentation masks as inputs. Several tools exemplify this approach: PathML is a python-based package that uses Skimage's "regionprops" method to calculate features such as volume, bounding boxes, and intensity. MCMICRO offers similar functionality through its MCQUANT method for Nextflow users [69]. Eng *et. al.* created a python-based package, cyclF_Validation, which focuses on improving processes for antibody specificity, signal removal, and batch normalization in cyclic immunofluorescence multiplex images [70]. Ilastik [66] uses non-deep learning machine learning, such as random forests, to count and track objects in addition to classification. Windhager *et al.* utilizes this latter and Cellprofiler in an end-to-end pipeline for multiplex images in R [71]. In addition, off-the-shelf software products also play a significant role in analyzing multiplex images, such as Inform® (Akoya Biosciences), which has historically been used for multispectral image analysis, fluorescent intensity quantitation, and rule-based phenotyping.

### 8.2.2 Deep Learning Applications (Pixel-Based Models)

Deep learning advancements in pathology have largely focused on image segmentation and classification. These areas of study have established a foundation for extensive research into image feature extraction and quantification, which are vital for constructing comprehensive end-to-end computational pathology pipelines. Many deep learning models, originally trained on various image types beyond multiplex-based images, demonstrate versatility in their applicability across different pathological image modalities. A significant feature of these models is their capability to process entire images as input, allowing for the extraction of features directly from the pixels, and eliminating the need for manual preprocessing of cell locations.

Feature extraction models can be grouped into two broad types: (i) models that create latent space features for downstream analysis and (ii) models that identify pertinent biological features. Both groups of models often rely on encoder-decoder networks commonly found in image segmentation models. Here we summarize the most recent applications: Dong *et al.* created an encoder-decoder network to learn latent space representations of image blocks for later tasks such as the classification of thyroid cancers [72]; Zong W. *et. al.* took a traditional approach by training a convolutional neural network (CNN) to create latent space representations of prostate multiparametric MRI (mpMRI) images which are then fed into a weighted extreme learning machine (wELM) classification algorithm [73]; MambaMIL model combines a feature extractor, linear projection, and a sequence reordering model to create features for downstream tasks such as survival prediction or cancer subtyping [74].

Creating models that can extract useful features can be difficult given data limitations in pathology. However, applying transfer learning and finetuning a model trained on general image data can yield useful results when building models for tissue classification and cross-cancer predictions for gene expression and mutation [75]. CLAM uses attention-based learning to process patches of a WSI into instance clusters [76].

Similar to feature extraction models, quantification deep learning methods also build on encoder-decoder methods found in segmentation research. Silina *et. al.* build on a typical encoder-decoder network (HookNet) to create Hooknet-TLS to quantify lymphoid aggregates in H&E images [77]. Liu *et al.* built off StarNet to create a model that quantifies myocardial inflammatory infiltration in H&E images [78]. In Haghighi *et. al.* researchers use ResNet-50 as the backbone of an encoder-decoder model to quantify dopamine neurons in Parkinson's Disease [79]. While much of the deep learning literature concerns H&E images, we provide the above examples as potential guides for future research on multiplexed images. Deep learning methodologies often can perform well across different areas of research and domain spaces.

### 8.3 Cell Annotation Methods in Multiplexed Imaging

Cell type annotation is a crucial step in understanding the cellular composition and the intricate dynamics within the tumor microenvironment (TME). In multiplexed imaging data, cell type annotation has traditionally depended on the expert knowledge of pathologists who manually identify, and label cell types based on morphological features and staining patterns. While this manual approach is accurate, it can be labor-intensive and subject to variability between observers. To improve the efficiency and reliability of cell phenotyping in multiplexed imaging data, automated or semi-automated annotation methods have gained prominence. These methods utilize well-characterized feature sets derived from known biomarker expressions and spatial distributions to identify cell types and states. Among the approaches employed are gating, unsupervised clustering, and the integration of deep learning techniques, all of which are increasingly acknowledged for their potential to transform cell phenotyping in multiplexed imaging.

### 8.3.1 Gating-based approaches

Gating is a technique adopted from flow cytometry and is among the most popular approaches for automated phenotyping in multiplexed imaging data. This approach employs a multi-dimensional space where each axis represents the intensity of a different fluorescent marker. Cells are selected or "gated" based on predefined intensity thresholds that correspond to known phenotypes. The spatial information can also be combined with the quantified fluorescence signals to enhance the accuracy of cell phenotyping. For instance, gating might be used to differentiate populations of CD8+ T cells from regulatory T cells by their respective marker expression profiles. This method's success hinges on the clear definition of phenotypes and the availability of markers that can reliably distinguish between them. However, gating's effectiveness can be reduced by technical noise factors, such as image processing artifacts or imperfect cell segmentation. While gating-based cell phenotyping does not specific computational tools, since it is entirely dependent on setting expert-defined decision rules, some publicly available tools offer gating-based functionality and intuitive user interfaces for cell type annotation. For instance, Cytomapper, an R toolkit for spatial data analysis, allows hierarchical gating based on the expression levels of up to 24 markers using a shiny interface [80].

### 8.3.2 Unsupervised clustering-based approaches.

Unsupervised clustering is another technique used to group cells based on phenotypic similarity, which is particularly useful for identifying novel cell types without prior bias [81–83]. This method is commonly employed in single-cell RNA-sequencing (scRNA-seq) experiments where cells are grouped into distinct clusters based on their gene expression profiles, with each cluster potentially representing a different cell type or state. Unlike gating, which relies heavily on the analyst's expertise and can introduce bias, unsupervised clustering offers a data-driven approach to cell-type identification. However, a benchmarking study by Hickey *et al.*, which compared hand gating to unsupervised clustering in annotating cell types in CODEX data, has shown that as the granularity of cell-type identification increases, the accuracy of labeling can decrease [84]. This poses a challenge for clustering algorithms which must balance the granularity of cell types with the confidence in the accuracy of their identification. The researchers managed this by avoiding overly subtle phenotype annotations during the clustering step, which can often lead to misclassification due to the continuous nature of marker expression levels. Moreover, the study recommended the use of over-clustering followed by spatial verification to refine cell-type identification [84]. In practice, both standalone clustering algorithms and those used in popular single-cell transcriptomics toolkits like Seurat [85,86] or Scanpy [87] can be employed for analyzing single-cell data from multiplexed imaging. Additionally, several publicly available tools have been developed specifically for analyzing data generated from spatially resolved technologies, incorporating clustering-based approaches for cell-type annotation. For example, Giotto, an R package, allows the use of various clustering algorithms such as Louvain or Leiden clustering on the single-cell data obtained from spatial transcriptomics or proteomics [88]. Another tool, ImaCytE, utilizes conventional dimensionality reduction and clustering methods for cell phenotyping in IMC data [89].

### 8.3.3 Semi-supervised clustering approaches.

Semi-supervised clustering has been used in the past for flow cytometry and mass cytometry data [90], but scarcely applied to mIF/IMC data. Recently, Seal *et al.* compared the performance of several algorithms including Random Forest (RF), Linear Discriminant Analysis (LDA) and Quadratic Discriminant Analysis (QDA) on different cohorts of mIF and IMC data, and found that RF showed better performances, compared to LDA and QDA [91].

## 9. Downstream Applications: From Data Analysis to Translational Implementation

Multiplexing technologies can be game-changing for many clinical tasks by enabling the simultaneous analysis and spatial mapping of multiple biological markers, such as predicting responses to immunotherapies across various cancer types. Immunotherapy has transformed cancer treatment, significantly improving outcomes in advanced cases [92]. However, these treatments are not universally effective, and their high cost and potential for severe side effects make it imperative to accurately predict which patients will benefit. Traditional methods, such as singleplex IHC for biomarkers like PD-L1, suffer from low reproducibility for several reasons, such as distinguishing PD-L1 expression between tumor cells and surrounding immune cells [28,93–95]. Additionally, these methods offer limited insights, as they fail to capture the full complexity of the TME. Multiplexed imaging techniques may overcome these limitations; for instance, the spatial proximity of PD-L1 and PD-1 positive cells in Merkel cell carcinoma and the proximity of cytotoxic

T cells to tumor cells in lung cancer are stronger predictors of immunotherapy response than PD-L1 expression alone [96,97]. Moreover, a recent meta-analysis, demonstrated how multiplex staining, allowing for detailed spatial analysis and co-expression profiling, outperforms other assays like gene expression profiling and tumor mutational burden in predicting immunotherapy response, across 10 solid tumor types [92].

Beyond the context of immunotherapy, the data obtained from cell segmentation and annotation in multiplexed imaging can be leveraged for various advanced analyses that have significant clinical implications such as the characterization of the TME and the spatial organization of immune, stromal, and tumor cells. In breast cancer, for instance, the spatial distribution of tumor-infiltrating lymphocytes (TILs) has been shown to correlate with better patient outcomes, making TILs density a valuable prognostic factor [98]. Furthermore, in ovarian cancer, studies using multiplex imaging have demonstrated that a higher density of intratumoral CD3+ T cells is predictive of improved survival [99].

Stromal-epithelial interactions are another key area of investigation, as these interactions are fundamental to cancer invasion and metastasis. Multiplexed imaging allows for detailed visualization of how cancer-associated fibroblasts (CAFs) and other stromal components interact with epithelial tumor cells, offering insights into mechanisms of treatment resistance. For example, the interaction between CAFs and epithelial cancer cells can drive resistance to chemotherapy in pancreatic cancer [100,101] and predict metastatic progression in localized prostate cancer [102]. By understanding these stromal-tumor dynamics, researchers can identify novel therapeutic targets to disrupt these interactions, potentially overcoming resistance.

Another promising application of multiplexed imaging is the neighborhood analysis, which quantitatively assesses the spatial relationships between different cell types within a tumor. CytoCommunity, an algorithm designed to identify cellular neighborhoods (CNs) based on cell phenotypes and their spatial distribution, analyzing risk-stratified colorectal and breast cancer data, revealed novel granulocyte- and CAFs-enriched CNs in high-risk tumors and uncovered altered interactions between neoplastic, immune, and stromal cells [103]. Using the CODEX® system, on FFPE tissue microarrays, Schürch *et al.* simultaneously profiled 140 tissue regions from 35 advanced-stage colorectal cancer patients using 56 protein markers, identifying 9 distinct CNs characteristics of the TME. Notably, the presence of PD-1+/CD4+ T cells within a granulocyte-rich CN was linked to improved survival. In contrast, disrupted inter-CNs communication was associated with poorer outcomes [6].

Multiplexed imaging is also increasingly being used to explore the heterogeneity of the TME across different cancer subtypes. For example, different brain malignancies, including primary gliomas and brain metastases, show distinct disease-specific immune landscapes in their TME [104,105].

As the field continues to evolve, multiplex imaging techniques are expected to play an even greater role in cancer diagnostics and therapeutics. By integrating multiplexed data with advanced computational tools such as machine learning and artificial intelligence, researchers and clinicians can uncover novel biomarkers, refine predictive models, and develop more precise and personalized treatment strategies for cancer patients. This approach has the potential to dramatically improve patient outcomes, especially in the era of precision oncology, where understanding the intricate spatial relationships within the TME is crucial for tailoring treatments to individual patients.

## 10. End-to-end Workflows for Multiplexed Image Analysis: Is It Feasible?

Analyzing multiplexed imaging data involves a complex, multistep process that requires the integration of various computational techniques to fully exploit the richness of the data. Establishing an end-to-end workflow presents significant challenges, chiefly the integration of multiple, often disparate, processing steps into a seamless pipeline (**Figure 2**). Each step—from image preprocessing and cell detection to feature extraction and phenotyping—requires a specialized approach. These steps must not only be effective in isolation but also harmonize with the entire pipeline to ensure the integrity and relevance of the data. The complexity of these workflows is further amplified by the diversity of multiplexed imaging modalities. Technologies such as IMC, MIBI, CODEX, mIF, and mIHC produce data with unique characteristics, necessitating workflow flexibility and adaptability. This is no small feat, considering that the workflow must maintain robustness across various file formats, imaging resolutions, and staining protocols. Furthermore, the sheer data volume generated by high-resolution multiplexed imaging poses a considerable challenge. The workflow must not only handle large datasets efficiently but also apply rigorous QC measures to detect and correct artifacts and inconsistencies, which could otherwise lead to inaccurate biological interpretations.

As the field of digital pathology evolves, an increasing number of tools have been developed to provide complete modular pipelines. These end-to-end workflows guide users from the acquisition of raw image data to the derivation of interpretable biological insights, by streamlining the analysis through the consolidation of essential steps, including image preprocessing, segmentation, feature extraction, and phenotyping, culminating in a structured output format (**Table 3**). This format is designed to be readily utilized for further statistical analysis or machine learning applications, facilitating a smoother transition from data to discovery.

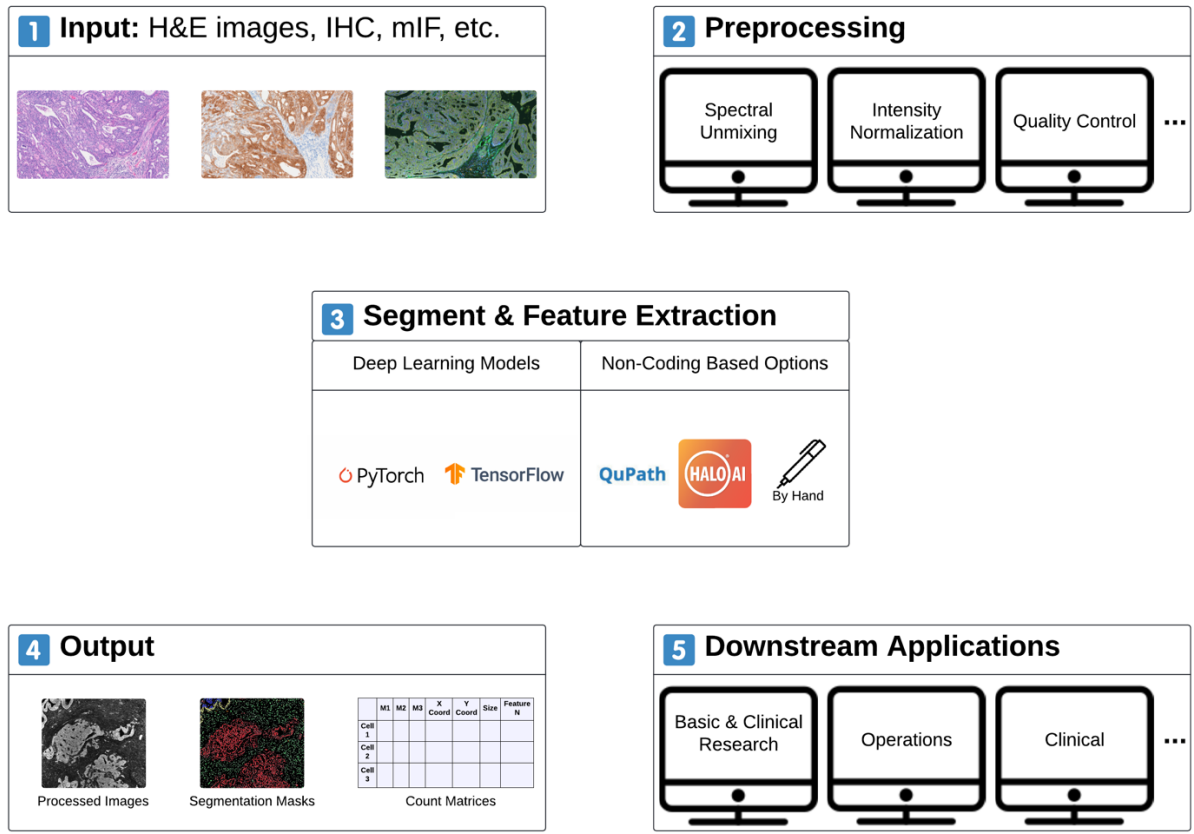

**Figure 2.** Overview of the steps of the analysis workflow for multiplex imaging modalities.

**Table 3.** List of publicly available end-to-end workflows for the analysis of multiplexed imaging data. mIF: multiplex immunofluorescence, mIHC: multiplex immunohistochemistry, IMC: imaging mass cytometry, MIBI: multiplexed ion beam imaging, CODEX: codetection by indexing

|  | **PathML** [11] | **Squidpy** [106] | **CytoKit** [107] | **MCMICRO** [69] | **SIMPLI** [108] | **Steinbock** [71] |
|---|---|---|---|---|---|---|
| ***Imaging Modalities*** | Various technologies including IMC, MIBI, CODEX, mIF, and mIHC | Various technologies including IMC, MIBI, CODEX, mIF, and mIHC | mIF | Various technologies including IMC, MIBI, CODEX, mIF, and mIHC | Various technologies including IMC, MIBI, CODEX, mIF, and mIHC | Mainly for IMC but can be modified to handle other modalities |
| ***Programming Language/Platform*** | Python | Python | python | Galaxy [109] and Nextflow [110] | | R |
| ***Input*** | Raw images | Raw images | Raw images | Raw images | Raw images | Raw images |
| *Stitching* | Supported using Groovy-based functionality | Not supported | Not supported | ASHLAR [111] | Not supported | Not supported |
| *Image Preprocessing* | Various preprocessing transforms including Gaussian, median, and box blur, normalization, superpixel interpolation, and morphological operations | Grayscale conversion and smoothing using Gaussian filter | Cycle registration and deconvolution | Illumination correction using BaSiC [112] | Includes normalization and background noise reduction capabilities | IMC preprocessing, including hot pixel filtering, denoising, and channel-to-channel spillover. Preprocessing other modalities is not supported |
| *Nuclei and Cells Segmentation* | Mesmer [61] or custom user-trained models | Watershed, StarDist [63,64] and Cellpose [62] | ClassifyPixels-Unet [113] | S3segmenter (watershed segmentation), UnMICST [67], Ilastic [66], Cypository | Deterministic and deep learning models (CellProfiler [60,113] and custom models) | Mesmer [61], Ilastic [66], CellProfiler [60,113], and Cellpose [62] |
| *Feature Extraction* | Segmentation feature extraction using Scikit-image | Segmentation feature extraction using Scikit-image | Segmentation feature extraction | Segmentation feature extraction using MCQuant [114] | Cell- and pixel-level feature extraction | Segmentation feature extraction using Scikit-image |
| *Cell phenotyping* | Gating and unsupervised clustering | Unsupervised clustering | Gating | FastPG [115]; unsupervised clustering approach derived from PhenoGraph | Gating and unsupervised clustering | Gating and unsupervised clustering using external packages |
| *Output Format for feature matrix* | AnnData object | AnnData object | FCS or CSV formats | AnnData object | Tabular files | AnnData object |

## 10.1 PathML: an all-in-one modular solution for advanced multiplexed image analysis

Among the aforementioned computational workflows for multiplexed image analysis, PathML stands out as a robust and comprehensive solution, which is designed to address the intricacies of the workflow challenges outlined previously [11]. Building upon the foundation of harmonizing various computational techniques, PathML excels in automating workflows that capitalize on the richness of data provided by diverse imaging technologies. With an emphasis on interoperability, PathML accommodates the nuances of different multiplexed imaging outputs, ensuring adaptability across various modalities with a standardized output format for feature matrix.

### 10.1.1 Enhanced Tile Stitching Capabilities

The digitization of pathology slides has ushered in an era where high-resolution images can reach sizes in gigabytes or even terabytes. For various preprocessing and image analysis tasks, these images need to be segmented into smaller, more manageable high-resolution image tiles. This is especially true for spectral unmixing and autofluorescence removal, employed for multiplexed images generated using the PhenoImager (Formerly known as Vectra Polaris) automated quantitative pathology imaging system (Akoya Biosciences) through the use of its native inForm software [4,116]. A crucial step before downstream analysis of spectrally unmixed images involves stitching these tiles together to reconstruct the original spatial context of the tissue. This reassembly is pivotal, as it restores the global perspective necessary for accurate pathological assessment and analysis.

Traditionally, software like QuPath [117] and HALO (Indica Labs) have offered robust solutions for tile stitching needs either through a GUI-enabled selection of tiled images or using custom Groovy scripts. However, PathML has recognized the imperative to integrate these essential tools into a unified, user-friendly framework. The PathML tile-stitching utility uses JPype, a Python Java interface [118], to meld the strengths of QuPath's stitching functionalities with PathML, facilitating a seamless stitching process without reliance on Groovy scripting, which is less familiar to a broad range of users. This new development in PathML is particularly adept at handling spectrally unmixed images, a common output from various multiplexed imaging techniques. The utility deftly processes a variety of TIFF inputs, converting them into standardized pyramidal OME-TIFF files—a format well-suited for downstream computational analyses. The incorporation of this Python-based stitching utility not only enhances PathML's capability but also widens the scope of its applicability, inviting an extensive user base that operates within Python-centric environments.

### 10.1.2 Streamlining Workflow Integration with GUI-Based AI Tools

The intersection of user-friendly graphical interfaces and the robust algorithmic capabilities of PathML epitomizes the convergence of accessibility and sophistication in digital pathology. PathML's strategic expansion to provide a seamless integration with GUI-based AI tools like QuPath [117] and HALO AI (Indica Labs) is a leap towards simplifying the adoption of AI in routine diagnostic practices. The integration through the Open Neural Network Exchange (ONNX) interface [119] is particularly notable since it allows pathologists to deploy HALO-trained models within the PathML ecosystem, thereby democratizing access to advanced computational tools for histopathology image analysis. This interface also mitigates the need for extensive computational resources or deep programming knowledge, which can often be barriers to technology adoption in clinical settings.

### 10.1.3 Inference API

To integrate seamlessly with previous PathML releases, the inference API is designed as a sequence of PathML transforms, so that each step can be slotted into existing PathML pipelines as a post-processing step. The inference classes are handled as PathML transforms. Thus, ONNX model predictions are saved as tiles to the SlideData object like how preprocessing transformations are saved. Representing the inference stages as PathML transforms also reduces the amount of code needed to complete an end-to-end analysis pipeline, as users can represent the entire pipeline from preprocessing to inference as a single list of PathML transforms.

### 10.1.4 Enabling Scalable Analysis Through Accessible AI Resources

The commitment of PathML to providing accessible AI tools is further exemplified by its Model Zoo, an ongoing development that hosts an array of pre-trained models, ready for deployment for diverse image analysis tasks. With these models at hand, PathML is set to reduce the barrier to entry for pathologists aiming to incorporate AI into their diagnostic and research workflows, regardless of their institutional capabilities. This model zoo is poised to become a dynamic repository of pre-trained models, hosted on the cloud through platforms like HuggingFace. This will allow users to access a multitude of state-of-the-art model architectures, fine-tuned for various pathology applications, with the simplicity of a single line of code initiation.

The models will be integral to the PathML inference API, facilitating seamless inference operations on image tiles and ensuring the results are systematically organized within the SlideData class—a structure conducive to subsequent analysis. In its current iteration, PathML offers a glimpse into the potential of this resource through the *RemoteTestHovernet*—a publicly available HoVer-Net model trained using the TIAToolbox library [120]. This proof-of-concept serves not only as a testament to the utility of the Model Zoo but also as a guide, with detailed example notebooks available to users for replicating and understanding the process.

### 10.1.5 Graph API: Delving into the Spatial Dynamics of Cellular Networks

The addition of the Graph API within PathML provides pathologists with a robust tool for studying the spatial architecture of cellular interactions. The structural and spatial organization of cells in a tissue are closely associated with their functional properties. This relationship has been extensively used in cellular and tissue imaging, facilitating research into the automated prediction of important biological functions from graphical representations of cells and tissues. In computational pathology, machine learning models that operate on cell and tissue graphs have demonstrated good efficacy in distinguishing between normal cellular formations and aberrant self-organizing clusters like those observed in cancer. Hence, in the latest version of PathML, we have incorporated popular methodologies for representing biological structures such as cells and tissues as graphs.

PathML's new Graph API provides methods for each step in the graph construction and analysis process. The first step involves image reading and stain normalization which can be done seamlessly using existing PathML workflows. Next, we provide a tissue detection method that is based on the Region Adjacency Graph construction process that iteratively merges nearby superpixels to identify tissue regions. For cell detection, we have traditional methods already implemented but one can also follow tutorials to use specialized deep learning models like HoVer-Net [121] using our 'Transforms' API. Feature extraction for each detected cell and tissue can be achieved using a newly implemented method that relies on pre-trained computer vision models like the ResNet. This process involves creating a patch around each cell or tissue, and then passing it into a pre-trained model to generate feature tensors. We provide graph construction classes that construct graphs from spatial coordinates of detected cells or tissues based on a K-nearest neighbor or a Region Adjacency method. These methods can also be extended by the user to implement other graph construction paradigms.

Additionally, we provide a framework for constructing machine learning models called Graph Neural Networks (GNNs) [122] that can leverage these graph structures for any downstream application. The features computed before can be used as node features during graph construction and are vital for training a GNN. The method is also flexible so that users can use their own models to generate the cell or tissue features. As an example, we provide a ready-to-use implementation of HACT-Net [123], a popular GNN, for users interested in training their own models. We also provide implementations of individual GNN layers so that other custom models can be implemented as well. These PathML graph-focused methods were recently used to predict outcome, resistance to immune checkpoint inhibitors, in a recent study in NSCLC [124].

### *10.1.6 Talk to PathML: a digital pathology assistant for democratizing access to advanced computational image analysis*

Lastly, leveraging the recent progress in medical Large Language Models (LLMs) [125], we created a new chat interface for those who would like to get started with PathML and have it suggested how to build pipelines from specific use-cases they might have. This was implemented by injecting all PathML examples and documentation into a Retrieval Augmented Generation (RAG) system based on GPT-4 capabilities [126]. Our "Digital Pathology Assistant" prototype, available here, can be used to build advanced end-to-end computational pipelines for specific use-cases [12]. Additionally, in the **S*upplementary Figure 1***, we report examples of how it can be used to generate specific computational pipelines for preprocessing and analyzing different types of imaging modalities.

## 11. Conclusion

New technologies mean new challenges and new opportunities and this innovative approach has significant challenges to face, one of the most significant lies in managing the exponential increase of datasets' size and simultaneous expansion of data dimensionality related to the complexity and density of data inferred for each case. On one hand, we have image size, which can range from relatively small files to extremely large datasets encompassing gigabytes of data due to high-resolution scans and multiplex staining platforms. On the other hand, we have image information density, which pertains to the amount of diagnostic, prognostic, and predictive information that can be extracted from these images. Managing a large number of cases necessitates robust cloud infrastructure and distributed computing capabilities to efficiently process and store vast datasets. Addressing high data dimensionality requires advanced spatial analysis methods to discern patterns across tissue structures and single-cell analysis to understand cellular-level variations. This trade-off between image size and information density is a pivotal issue in computational pathology, necessitating advanced solutions for efficient image compression, data management, and analytical techniques to maximize diagnostic utility while minimizing resource consumption. Moreover, effective visualization tools are essential to handle data scales ranging from gigabytes to petabytes, enabling researchers and clinicians to interpret complex data intuitively and accurately. To maximize innovation and patient impact, digital and computational pathology strategies must scale in both dimensions: the dataset size and the dimensionality of data.

To effectively support these research efforts, software tools must be designed with a strong emphasis on scalability, standardization, and ease of use. PathML, an open-source framework crafted with these best practices at its core, aims to lower the barrier to entry for digital pathology. It enables researchers to implement a variety of computational pathology use cases with just a few lines of code [11]. It offers comprehensive support for all facets of computational pathology research and includes capabilities for loading a wide range of imaging modalities and file formats, constructing modular and fully customizable preprocessing pipelines, and leveraging parallel computing. Additionally, it integrates seamlessly with other tools in machine learning, AI, and single-cell analysis ecosystems becoming a versatile and powerful tool to address numerous biologically relevant problems.

## Acknowledgments

Authors are grateful to the many colleagues who contributed the insightful conversations, especially those in the Department of Pathology and Laboratory Medicine of Weill Cornell Medicine and the Department of Informatics and Analytics of Dana-Farber Cancer Institute.

## Disclosures

The authors declare no competing interests. M.O. and L.M. are supported by the NCI grant U54CA273956. G.N.F. is supported by a fellowship from the American Italian Cancer Foundation (AICF) and by the Italian Ministry of University and Research - PON "Research and Innovation" 2014-2020 (PON R&I) Actions IV. 4 "Doctorates and research contracts on innovation topics". M.L. is supported by the National Cancer Institute (NCI) grants P50CA211024 and P01CA265768, the USA Department of Defense (DoD) grant DoD PC160357, as well as the Prostate Cancer Foundation.

## Author contribution

First draft: MO, GNF, FS, JW, SR, VU, AC, IV, CS, and RU; digital pathology analyses: MO, GNF, and FS; software implementation: MO, VU, SR, JW, AC, RU; supervision: LM, RU, ML. All authors have read and agreed to the final draft of the manuscript.

## Other disclosures

GPT4DFCI, a private and secure generative AI tool based on GPT-4 models and deployed at Dana-Farber Cancer Institute for non-clinical use, was used to improve this manuscript: it was used to reword certain passages and further critique our work.

## Code and data availability

The open-source version of PathML is available under GNU GPL v2 license at https://github.com/Dana-Farber-AIOS/pathml. Commercial licenses are also available. The instructions and knowledge base that power our Digital Pathology Assistant v3.0 are also available at https://github.com/Dana-Farber-AIOS/pathml. The Digital Pathology Assistant v3.0 is available at https://chat.openai.com/g/g-L1IbnIIVt-digital-pathology-assistant-v3-0 for all OpenAI ChatGPT Pro users to try. The de-identified pathology images used in this manuscript are available from the corresponding authors upon reasonable request.

**Supplementary Table 1. Glossary of key terms.** Since artificial intelligence is such a fast-moving target and quickly evolving topic, readers interested in specific and new definitions should consult the Wikipedia Glossary of AI (https://en.wikipedia.org/wiki/Glossary_of_artificial_intelligence)

| | |
|---|---|
| **Annotation** | Indication of the position and/or outline of structures or objects within digital images. They may include labels and other metadata. They can be manually generated or produced using algorithmic tools. |
| **Artificial Intelligence (AI)** | A field of computer science that deals with the development and study of programs that enable machines to learn so that they can achieve a well-defined goal. |
| **Artificial Neural Network (ANN)** | A type of machine learning model inspired by the neuronal structure and function in the brain that enables it to learn from input data to achieve a well-defined goal. |
| **Computational pathology** | A field that applies computer-based technology and algorithms to analyze digital images derived from pathology specimens, such as tissue sections, to enhance diagnostic accuracy and prediction of response to therapy. |
| **Contrast enhancement** | Manipulate the intensity of the individual pixels in a digital image to increase visibility |
| **Convolutional Neural Network (CNN)** | A type of ANN commonly applied to visual images. |
| **Diffraction limit** | The limit on an optical system's ability to resolve detail, determined by the wavelength of light and the system's numerical aperture, which restricts how closely two points can be distinguished. |
| **Digital Pathology** | Acquisition, management, sharing, and interpretation of pathology information/slides in a digital environment (on any digital device) |
| **Data Augmentation** | The process of artificially increasing the size of your dataset by applying modifications or transformations on the existing data points. |
| **Data matrices** | Structured formats used to organize and analyze complex data sets. In pathology, these often represent quantified features from image analysis, such as cell counts or marker intensities. |
| **Deep Learning** | A subset of machine learning that uses algorithms modeled loosely after the human brain (neural networks) to learn from large amounts of data. It is particularly effective for tasks that involve image recognition, speech recognition, and language processing, which involve recognizing patterns and making decisions based on large datasets. |
| **Encoder-Decoder Networks** | A type of neural network used in deep learning for tasks that involve converting the format or type of input data into a desired output format, commonly used for image segmentation tasks. |
| **Features extraction** | The technique of identifying and quantifying specific characteristics from segmented images, such as shape, size, |

| | or texture, can be used for further statistical analysis. Data can be summarized in count matrices |
|---|---|
| **Gating** | A method adapted from flow cytometry used in multiplexed imaging to categorize cells based on the intensity of fluorescence markers. It involves setting thresholds that define cell types or states. |
| **Graphical Representations of Cellular Interactions** | Visual models that map the relationships and interactions between different cell types within a tissue, offering insights into cellular behavior and tissue architecture. |
| **Graph Neural Network (GNN)** | Neural network that works on graph-structured data, learning patterns from nodes and their connections to solve tasks like predicting relationships or classifying nodes. |
| **Grayscaling** | Convert a color image to shades of gray, removing all color information while preserving the brightness levels, resulting in an image where each pixel represents a shade from black to white. |
| **Ground truth** | Information that is known to be real and true, often generated using observations or measurements. In some types of machine learning, called supervised machine learning, it is required so that the model can aim for this value in the learning process. |
| **In situ proteomics and spatial transcriptomics** | Advanced methods that map protein and gene expression directly within tissue sections, preserving spatial relationships and providing context to molecular data. |
| **k-nearest neighbor** | A supervised machine learning algorithm that makes predictions based on the proximity of data points. |
| **Image binarization** | converts the color or grayscale images into digital binary images consisting of small black and white pixels to make classifier algorithms more efficient. |
| **Large language model (LLM)** | AI model trained on massive text data to understand and generate human-like language, enabling it to perform tasks like answering questions, summarizing text, and engaging in conversation. |
| **Machine Learning (ML)** | A branch of artificial intelligence that uses data and well-defined goals to help computer programs learn to fulfill a desired task. |
| **Metadata** | Data elements that accompany a data point. In image data, the metadata can refer to the type of image, the date it was generated, and so on. |
| **Multiplexed imaging** | A technology that allows the simultaneous detection of multiple biomarkers in a single tissue section. This technique provides a comprehensive view of the tissue's cellular and molecular composition. |
| **Open Neural Network Exchange (ONNX)** | An open-source format for representing machine learning models, enabling models to be transferred across different frameworks, for easier deployment and interoperability |
| **Phenotyping** | The process of classifying cells or tissues based on their observable characteristics or measured features, often using automated or semi-automated methods. |
| **Photobleaching** | Irreversible loss of fluorescence, after the prolonged light exposure |
| **Photodamage** | Damage to cells or tissues caused by exposure to intense light, which can impair cell function or cause cell death. |
| **Pixel Normalization** | The process of scaling pixel values in an image to a standard range (e.g., 0 to 1) to improve model training stability and performance in computer vision tasks. |
| **Preprocessing** | The initial steps in data analysis that prepare images for further examination. This may include enhancing image quality, correcting distortions, and removing artifacts. |
| **Pyramidal Image** | Multi-scale image representation that involves different stages where each stage views the image at a different resolution. |
| **Quantitative histopathology** | A branch of pathology that uses quantitative methods to measure and analyze histological features from tissue samples. This approach leverages digital imaging and computational tools to objectively assess characteristics such as cell count, marker intensity, and morphological structures, providing a more precise and reproducible analysis than traditional qualitative manual methods. |
| **Random forest** | A type of machine learning model that combines the output of multiple decision trees to perform predictions. |
| **Recurrent neural networks (RNN)** | A type of deep learning architecture designed for sequential data like text or time series. |
| **Residual neural network (ResNet)** | A type of deep learning architecture that leverages residual connections between neurons in each layer to improve information extraction at each layer. |
| **Resizing** | Adjust an image's dimensions to fit a specific scale or resolution. |
| **Segmentation** | The process of dividing an image into parts to isolate regions of interest, such as individual cells or specific tissue components. This is crucial for detailed analysis. |
| **Semi-supervised clustering** | An approach that integrates the given prior information (e.g., class labels and pairwise constraints) into clustering to guide the clustering process and improve the performance. |
| **Spectral Unmixing** | A computational technique used in multiplexed imaging to separate overlapping signals of different fluorophores in a specimen. This method corrects for 'bleed-through' between channels, where the emission spectrum of one fluorophore overlaps with another, ensuring accurate quantification of each fluorescent signal. |
| **Spectral bleed-through** | Occurs when fluorescence signals from different dyes overlap in imaging, causing accidental mixing of signals between channels and reducing image clarity. |

| Tiling | Divide an image into smaller segments (tiles) to facilitate analysis of high-resolution data in manageable sections. |
|---|---|
| **Unsupervised clustering** | A machine learning approach that groups data (e.g., cells) based on their similarities without prior labeling, useful in discovering new cell types or patterns in data. |
| **Whole slide image (WSI)** | A digitized histopathology glass slide created on a slide scanner. The digitized glass slide represents a high-resolution replica of the original glass that can then be manipulated through software to mimic microscope review and diagnosis. Also referred to as a virtual slide. |

**Supplementary Table 2.** List of common tools for multispectral unmixing.

| | Traditional Linear Unmixing | PICASSO [1] | Maric *et al.* [2] | LUMoS [3] | NMF [4] |
|---|---|---|---|---|---|
| *Algorithm/Method* | Unmixing based on linear decomposition of the different wavelengths | Mutual information separation based on iterative subtraction of scaled images | Semi-supervised model with bleed-through estimation done with LASSO regression | Machine learning based clustering | Non-negative Matrix Factorization |
| *Image dimensions* | 2D, 3D | 2D, 3D | 2D, 3D | 2D, 3D | 2D, 3D |
| *Assumptions* | Presence of reference spectral library | Mixed images show "mutual information" (spectral mixing) that can be minimized | The signal of interest is always brighter than the bleed-through | Each pixel represents a unique fluorophore (problem for co-localization) | Almost every pixel has a channel with no fluorophore emission |
| *Key Advantages* | Works even for very low signal | Easy to deploy, no reference library needed | No reference library | Fast unmixing, useful if N fluorophores > N detection channels | Low computational cost, robust results |
| *Potential Pitfalls* | Need for spectral library | Not tested by other groups | User needs to indicate which channels are to be unmixed, not tested enough | Need to modulate parameters, untested on widefield fluorescence slides | Poor performance reported on another paper [5,6]. |
| *Need for spectral library reference* | Yes | No | No | No | No |
| *Language used* | Undisclosed | MATLAB | Python | Python | Python |
| *License* | Commercial license (Inform, Akoya Bioscience) | Open Source (Supplementary software in link) | Open Source | Open Source | Open Source |

**Supplementary Table 3.** List of multiplexed imaging datasets with available nuclear and/or cell annotations; Only expert-annotated multiplexed imaging datasets were considered in this table. MIBI: multiplexed ion beam imaging; CyCIF: cyclic immunofluorescence; CODEX: codetection by indexing; IF: immunofluorescence; LUAD: lung adenocarcinoma; SCLC: small-cell lung cancer; PDAC: pancreatic ductal adenocarcinoma; COAC: colon adenocarcinoma; CTCL: cutaneous T-cell lymphoma; BCC: basal cell carcinoma; SCC: squamous cell carcinoma

| | TissueNet [1] | Aleynick et al. [2] | Kromp et al. [3] | DeepLIif [4] | Han et al. [5] |
|---|---|---|---|---|---|
| **Number of images** | 3850 | 211 | 79 | 1667 | 223 |
| **Imaging modalities/platforms** | Vectra, MIBI, CODEX, and CyCIF | Sequential IF with unmixing (Akoya Vectra 3.0), sequential IF with narrowband capture (via Ultivue InSituPlex with Zeiss Axioscan image capture) and cyclical IF with narrowband capture (via Akoya CODEX) | Zeiss Axioplan II, Zeiss- and Leica laser scanning microscopes (LSM) | Three-color IF, DAPI only staining, chromogenic IHC and HE acquired with Axioscan Zeiss | Multiplex IF from GE Research for human, cyclic IF MELC for mouse[58] |
| **Tissue type** | Pancreas, breast, tonsil, colon, lymph node, lung, esophagus, spleen, and skin | Lung (LUAD and SCLC) , breast (Paget's disease and ductal carcinoma) pancreas (PDAC), colon (COAC), Lymph nodes (normal, Hodgkin's lymphoma and Merkel cell carcinoma), tonsils (normal), ovary (ovarian serous carcinoma), skin | Cryosections from ganglioneuroblastoma, Wilms tumor, neuroblastoma, bone marrow cytospin preparations infiltrated with neuroblastoma cells, neuroblastoma tumor touch imprints, cells of two neuroblastoma cell lines (CLB-Ma, STA-NB10) cytospinned on microscopy glass slides and cells of a normal human keratinocyte cell line (HaCaT) cytospinned or | FFPE samples of bladder carcinoma and non-small cell lung carcinoma stained for Lap2-beta, panCK and KI67. | |

| | | (extramammary Paget's disease, CTCL, and BCC), tongue (SCC) | grown on microscopy glass slides | | |
|---|---|---|---|---|---|
| ***Species*** | Human, Mouse, and macaque | Human | Human | Human | |
| ***Marker(s)*** | DAPI + membrane or cytoplasm marker | +40 markers including DAPI | DAPI | Hoechst, Lap2-beta | |
| ***Available Annotations*** | Nuclei and cells | Nuclei and cells | Nuclei | Nuclei | |

**Supplementary Figure 1.** Digital Pathology Assistant v3.0 (available here) homepage and examples of how our AI can be used to build complex analysis pipelines, by simply asking the AI. This example leverages OpenAI GPT-4 and its retrieval-augmented generation capabilities to translate user questions into executable PathML code.

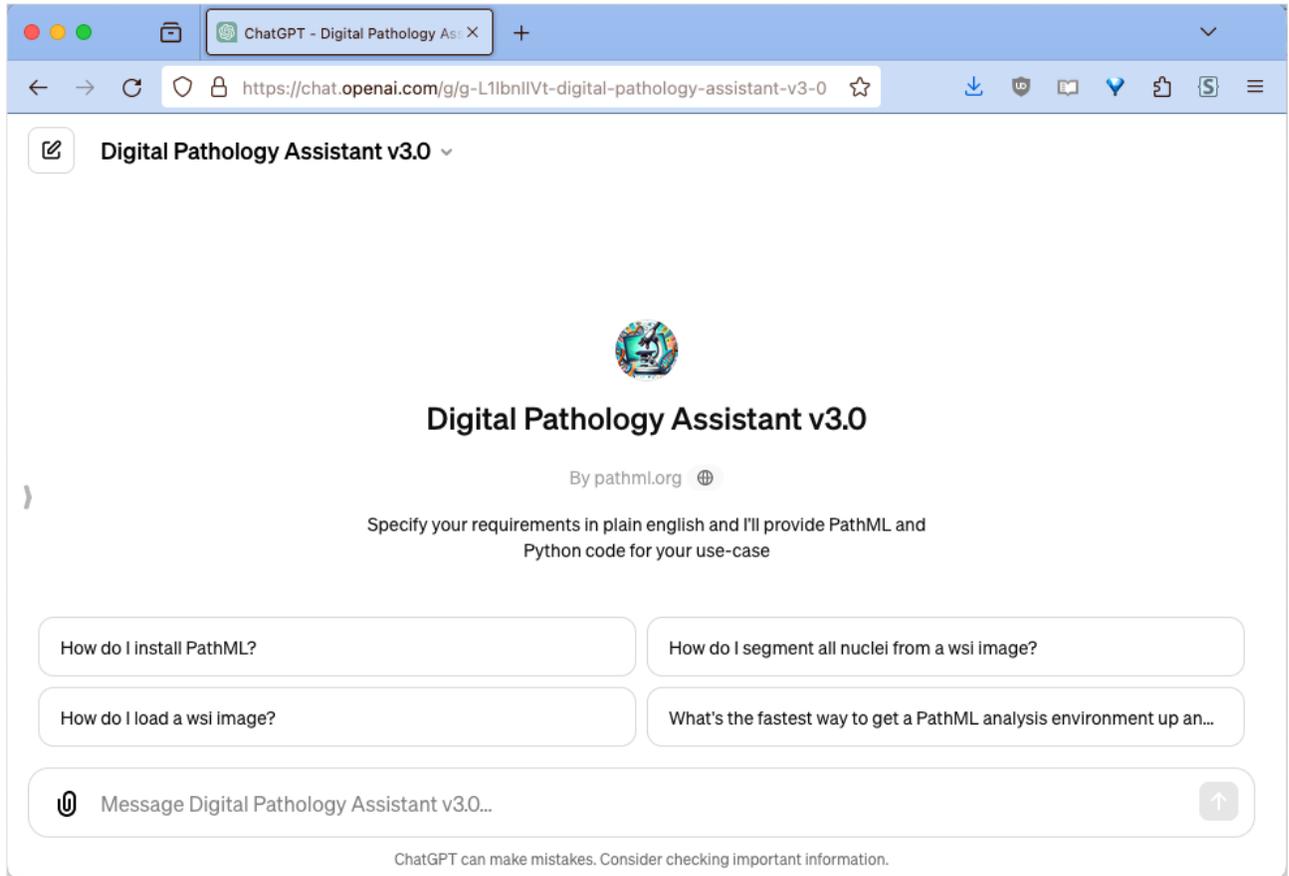

**You**

How do I quantify the intensity of a nuclear marker in a Vectra Polaris image?

**Digital Pathology Assistant v3.0**

🧑‍🔬🔬 Digital Pathology Assistant v3 Initiated 🎓.

Quantifying the intensity of a nuclear marker in a Vectra Polaris image using PathML involves several steps, from image loading and preprocessing to segmentation and quantification. Here's a general approach to accomplish this:

### Step 1: Load the Vectra Polaris Image

First, import the necessary class from PathML and load your Vectra Polaris slide. If your image is a multiparametric immunofluorescence (mIF) image, you might use the `VectraSlide` class.

```python
from pathml.core import VectraSlide

slide = VectraSlide(filepath="path_to_your_vectra_image")
```

### Step 2: Preprocess the Image

Preprocess the slide to prepare it for analysis. For Vectra Polaris images, this often includes collapsing runs to reduce the dimensionality of the image data.

```python
from pathml.preprocessing import Pipeline
from pathml.preprocessing.transforms import CollapseRunsVectra

pipeline = Pipeline([
    CollapseRunsVectra()
])

slide.run(pipeline)
```

### Step 3: Segment Nuclei

Segment the nuclei within the slide. For this, you can use a segmentation model suitable for your image type. PathML doesn't directly provide a built-in nuclear segmentation model, but you can integrate it with models from other libraries (e.g., using a pre-trained deep learning model from PyTorch or TensorFlow) or apply custom segmentation approaches.

```python
# Example placeholder for segmentation step
# This step will vary significantly depending on the chosen method and model
segmented_nuclei = custom_nuclei_segmentation(slide)
```

### Step 4: Quantify Nuclear Marker Intensity

After segmentation, you'll have regions corresponding to nuclei. The next step is to quantify the intensity of your marker of interest within these regions. This process will depend on the output of your segmentation step and might involve calculating the mean, median, or sum of pixel intensities within the nuclear regions for the specific channel of your marker.

```python
# Example pseudo-code for quantification
marker_intensities = quantify_nuclear_marker_intensity(segmented_nuclei, channel="mar
```

This step requires you to define or use a function like `quantify_nuclear_marker_intensity`, which calculates the intensity of your nuclear marker within the segmented nuclei. The implementation of this function will depend on how your nuclear regions are represented (e.g., masks, coordinates) and how you wish to measure intensity (e.g., mean intensity per nucleus).

### Note

- The specifics of segmentation and intensity quantification will depend on the details of your marker, the quality of your images, and the requirements of your analysis.
- Integration with deep learning models or custom algorithms might require additional libraries such as `torch`, `tensorflow`, or `scikit-image`.

🎓...more details can be found at https://pathml.readthedocs.io/en/latest/search.html?q=Vectra+Polaris+analysis

**You**

How do I use PathML Graph API to construct cell graphs?

**Digital Pathology Assistant v3.0**

🧑‍⚕️ Digital Pathology Assistant v3 Initiated 🎓.

To construct cell graphs using PathML's Graph API, you can follow these steps:

1. Load your slide image into a `SlideData` object, specifying the slide type if known.
2. Define a preprocessing pipeline with necessary transformations using the `Pipeline` class.
3. Use a graph builder from `pathml.graph`, such as `KNNGraphBuilder`, to create the graph.

Here's an example of how you might write the code:

```python
from pathml.core import HESlide
from pathml.preprocessing import Pipeline, NucleusDetectionHE
from pathml.graph import KNNGraphBuilder
from pathml.graph.utils import get_full_instance_map

# Define the path to your slide
slide_path = 'path_to_your_slide'

# Initialize your slide
wsi = HESlide(slide_path, name="Your Slide Name", backend="openslide", stain='HE')

# Define your preprocessing pipeline
pipeline = Pipeline([
    NucleusDetectionHE(mask_name="nuclei")
])

# Run the pipeline on the whole-slide image
wsi.run(pipeline)

# Get the nuclei segmentation masks and centroids
image, nuclei_map, nuclei_centroid = get_full_instance_map(wsi, mask_name="nuclei")

# Initialize the graph builder
knn_graph_builder = KNNGraphBuilder(k=5, thresh=50, add_loc_feats=True)

# Build the cell graph
cell_graph = knn_graph_builder.process(nuclei_map, return_networkx=True)
```

This code snippet does the following:

- Loads a whole-slide image (`HESlide` for H&E stained images).
- Sets up a preprocessing pipeline for nuclei detection.
- Applies the pipeline to detect nuclei in the whole-slide image.
- Extracts the nuclei segmentation masks and centroids.
- Uses `KNNGraphBuilder` to construct a cell graph based on the detected nuclei.

Make sure to adjust the parameters like `slide_path`, `name`, and any pipeline steps according to your specific use case and slide type.

🎓 …more details can be found at https://pathml.readthedocs.io/en/latest/search.html?q=graph+construction

**You**

how do I create a mIF analysis?

**Digital Pathology Assistant v3.0**

🤖 📚 Digital Pathology Assistant v3 Initiated 🎓.

To conduct a multiparametric immunofluorescence (mIF) analysis using PathML, follow these general steps. The process involves reading the image, preprocessing it (which may include collapsing runs for certain platforms), segmenting the image, and then analyzing the segments. The specific steps can vary based on the image platform (e.g., Vectra Polaris, CODEX).

### Step 1: Load the mIF Slide

First, import the necessary classes and load your mIF slide. Depending on your image source, you might use `VectraSlide`, `CODEXSlide`, or `MultiparametricSlide`.

```python
from pathml.core import CODEXSlide

# Adjust the file path to point to your mIF slide
slide_path = "path_to_your_mif_image"
slide = CODEXSlide(filepath=slide_path)
```

### Step 2: Preprocess the Slide

Before segmentation, you may need to preprocess the slide. For Vectra and CODEX slides, this often involves collapsing runs.

```python
from pathml.preprocessing import Pipeline
from pathml.preprocessing.transforms import CollapseRunsCODEX

# Example for CODEX slides
pipeline = Pipeline([
    CollapseRunsCODEX()
])

slide.run(pipeline)
```

### Step 3: Segment the Slide

Use a segmentation model appropriate for mIF images. For instance, `SegmentMIF` based on the Mesmer model is often used for segmenting cells in mIF images.

```python
from pathml.models import SegmentMIF

# This is a simplified example. You might need to adjust parameters or the model based
model = SegmentMIF()

# Segment the slide
segmented_slide = model.apply(slide)
```

### Step 4: Analyze the Segmentation

After segmentation, you can analyze the segments to extract meaningful biological or clinical information. This might involve quantifying marker expression per cell, clustering cells based on marker expression, or other analyses specific to your research question.

The details of this step are highly specific to the analysis you wish to perform and the data you are working with.

```python
# Example pseudo-code for analysis
results = perform_custom_analysis(segmented_slide)
```

Replace `perform_custom_analysis` with your analysis function, which could involve operations like quantification, clustering, or spatial analysis.

### Note

- The code provided here is a simplified outline. Details like file paths, specific preprocessing steps, model parameters, and analysis functions need to be tailored to your specific dataset and research questions.
- Ensure all necessary components (e.g., models for segmentation) are properly configured in your environment.

🎓...more details can be found at https://pathml.readthedocs.io/en/latest/search.html?q=mIF+analysis

**You**

How do I use the PathML Inference API to programmatically run an ONNX model I exported?

**Digital Pathology Assistant v3.0**

🧑‍⚕️🔬 Digital Pathology Assistant v3 Initiated 🎓

To use the PathML Inference API for running an ONNX model programmatically, follow these steps. This process involves setting up the inference environment, loading your slide, preparing the model if necessary, and executing the model using PathML's Inference API.

### Step 1: Install ONNX Runtime

Ensure you have the `onnxruntime` package installed in your environment, as it's required for running ONNX models. If it's not installed, you can add it via pip:

```bash
pip install onnxruntime
```

### Step 2: Load Your Slide

First, load the slide you want to analyze. Depending on the slide type, you might use `HESlide`, `VectraSlide`, `CODEXSlide`, or the generic `SlideData`:

```python
from pathml.core import HESlide

slide = HESlide(filepath="path/to/your/slide.svs")
```

### Step 3: Preprocess Your ONNX Model (If Necessary)

Before using your model for inference, make sure it's properly formatted and compatible with ONNX Runtime. Sometimes, you might need to preprocess your ONNX model to remove initializers from its inputs:

```python
from pathml.inference import remove_initializer_from_input

original_model_path = "path/to/your/model.onnx"
fixed_model_path = "path/to/your/fixed_model.onnx"

# This step is optional and depends on your model's requirements
remove_initializer_from_input(model_path=original_model_path, new_model_path=fixed_mo
```

### Step 4: Configure the Inference

Set up an `Inference` object with your ONNX model. You'll need to specify the path to the model, the name of the input layer, the number of classes (for classification tasks), and the type of model (e.g., segmentation or classification):

```python
from pathml.inference import Inference

inference = Inference(
    model_path=fixed_model_path,
    input_name="input",  # Adjust this to match your model's input layer name
    num_classes=2,  # Set this based on your model's output
    model_type="segmentation"  # Use "classification" or "segmentation" as appropriat
)
```

### Step 5: Run the Inference Pipeline

Incorporate the inference process into a PathML pipeline and execute it on your slide:

```python
from pathml.preprocessing import Pipeline

pipeline = Pipeline([inference])

# Execute the pipeline on your slide
# Adjust parameters like tile_size and level as necessary for your model and slide
slide.run(pipeline, tile_size=256, level=0)
```

### Additional Notes:

- Make sure the `tile_size` and `level` parameters in `slide.run()` are set according to your model's input size requirements and the desired resolution for analysis.
- The process might vary slightly depending on specific requirements of your ONNX model or the preprocessing needs of your slides.

🎓...more details can be found at https://pathml.readthedocs.io/en/latest/search.html?q=inference+API

**You**

What kind of images can I analyze in pathml?

**Digital Pathology Assistant v3.0**

🔬🖥️ Digital Pathology Assistant v3 Initiated 🎓.

In PathML, you can analyze a wide variety of image types relevant to digital pathology and computational pathology research. The types of images you can work with include, but are not limited to:

1.

**Hematoxylin and Eosin (H&E) Stained Slides**: The most common type of slide used in pathology, stained to highlight the cellular and structural components of tissue. PathML offers specialized support for these slides through the `HESlide` class.

2.

**Immunohistochemistry (IHC) Stained Slides**: Used to detect specific antigens in cells of a tissue section by exploiting the principle of antibodies binding specifically to antigens in biological tissues.

3.

**Multiparametric Immunofluorescence (mIF) Slides**: These slides are processed through techniques like multiplex immunofluorescence that allow for the visualization of multiple targets or markers within the same sample, providing rich information on the spatial distribution and co-localization of proteins. PathML supports multiparametric slides through classes like `MultiparametricSlide`, `VectraSlide` (for Vectra Polaris images), and `CODEXSlide` (for CODEX platform images).

4.

**Fluorescence In Situ Hybridization (FISH) Slides**: A technique used to detect and localize the presence or absence of specific DNA sequences on chromosomes.

5.

**Whole-Slide Images (WSIs) from Various Scanners**: PathML supports WSIs in formats produced by different slide scanners, thanks to its integration with OpenSlide, BioFormats, and support for DICOM files.

**You**

How do I install PathML on a mac ?

**Digital Pathology Assistant v3.0**

🔬🖥️ Digital Pathology Assistant v3 Initiated 🎓.

To install PathML on a Mac, you can follow these simplified steps:

1. **Install Conda** (Recommendation: use Miniconda for a lightweight installation).
2. **Update Conda and Optionally Use libmamba** for a faster dependency resolution (optional step but recommended).
3. **Install External Dependencies**:
- OpenSlide using Brew: `brew install openslide`.
- OpenJDK: `conda install -c conda-forge 'openjdk<=18.0'`.

4. **Create and Activate a Conda Environment**:
- Create the environment: `conda create --name pathml python=3.9`.
- Activate the environment: `conda activate pathml`.

5. **Install PathML** using pip: `pip install pathml`.

After following these steps, you should have PathML installed and ready for use on your Mac.

For further details and updates on the installation process, please refer to the PathML documentation.

🎓...more details can be found at [https://pathml.readthedocs.io/en/latest/search.html?q=installation](https://pathml.readthedocs.io/en/latest/search.html?q=installation)